\documentclass[11pt,draftcls,onecolumn]{IEEEtran}

\usepackage{amsmath,amsfonts,epsfig,graphicx,color,multirow,subfigure,textcomp}

\begin{document}

\title{Optimal Energy Consumption Model for \\ Smart Grid Households with Energy Storage}

\author{Jayaprakash~Rajasekharan,~\IEEEmembership{Student,~IEEE,} and~Visa~Koivunen,~\IEEEmembership{Fellow,~IEEE}

\thanks{J. Rajasekharan and V. Koivunen are with SMARAD CoE, Department of Signal Processing and Acoustics, Aalto University, P.O.Box 13000, FI-00076, Espoo, Finland. e-mail: \{jayaprakash.rajasekharan,visa.koivunen\}@aalto.fi.}

\thanks{Some preliminary ideas and results have been submitted to IEEE ICASSP 2014 conference.}}

\maketitle

\begin{abstract}

In this paper, we propose to model the energy consumption of smart grid households with energy storage systems as an intertemporal trading economy. Intertemporal trade refers to transaction of goods across time when an agent, at any time, is faced with the option of consuming or saving with the aim of using the savings in the future or spending the savings from the past. Smart homes define optimal consumption as either balancing/leveling consumption such that the utility company is presented with a uniform demand or as minimizing consumption costs by storing energy during off-peak time periods when prices are lower and use the stored energy during peak time periods when prices are higher. Due to the varying nature of energy requirements of household and market energy prices over different time periods in a day, households face a trade-off between consuming to meet their current energy requirements and/or storing energy for future consumption and/or spending energy stored in the past. These trade-offs or consumption preferences of the household are modeled as utility functions using consumer theory. We introduce two different utility functions, one for cost minimization and another for consumption balancing/leveling, that are maximized subject to respective budget, consumption, storage and savings constraints to solve for the optimum consumption profile. The optimization problem of a household with energy storage is formulated as a geometric program for consumption balancing/leveling, while cost minimization is formulated as a linear programming problem. Simulation results show that the proposed model achieves extremely low peak to average ratio in the consumption balancing/leveling scheme with about 8\% reduction in consumption costs and the least possible amount for electricity bill with about 12\% reduction in consumption costs in the cost minimization scheme.

\end{abstract}

\begin{IEEEkeywords}
Smart grids, energy storage, intertemporal trade, optimal consumption, geometric programming.
\end{IEEEkeywords}

\IEEEpeerreviewmaketitle

\section{Introduction}

Energy storage has recently gained attention due to the integration of fluctuating and intermittent renewable energy sources and plug-in hybrid electric vehicles (PHEVs) into smart grid systems \cite{smartgrid}. In addition to being served by utility companies, households connected to the smart grid may privately own and operate renewable energy sources such as solar panels, wind turbines, etc. along with energy storage systems (ESS). Utility companies deploy multiple sensors to help them monitor, study, evaluate and meet the demand generated by households throughout the distribution network. Demand side management (DSM) commonly refers to programs employed by the utility company that control the energy consumption of households. DSM programs such as residential load management aim at reducing and/or shifting/scheduling consumption at the household level to off-peak periods by means of smart pricing. Examples of such pricing options are critical peak pricing (CPP) \cite{20072121}, time-of-use pricing (ToUP) \cite{6266720}, real time pricing (RTP) \cite{6547835}, etc. Smart pricing combined with fluctuating renewable energy production make energy consumption schedulers (ECS) and energy storage systems \cite{5582228} indispensable in smart homes. Though scheduling \cite{5628271} itself is successful to a certain extent in reducing the peak to average ratio (PAR) of consumption and yielding some cost savings, there is a limit to the amount of household energy requirements that can be scheduled wihtout causing excessive discomfort. Even though scheduling is implicitly supported by the utility company through smart pricing dynamics, scheduling cannot guarantee absolute consumption cost minimization for the household or consumption PAR minimization for the utility company. However, energy storage systems provide smart homes with an attractive option to either minimize their consumption costs or level their consumption such that they can present the utility company with a demand that is as uniform as possible.

Residential energy storage is enabled by dedicated battery systems, supercapacitors, PHEVs, etc. \cite{6304671}. Vehicle to home (V2H) and vehicle to grid (V2G) technologies \cite{6571224} have already enabled bidirectional transfer of energy between the grid or home and the battery system in PHEVs with the aim of selling demand response services back to the grid or home. Though energy storage using batteries has been traditionally considered lossy, difficult and expensive, it is expected to be a key component of smart homes in future smart grid systems. Home battery back-up systems with storage capacities around 2-5 KWh have been in existence for quite some time and their prices have seen a steady decline due to wide spread adoption along with renewable energy production sources. Toshiba has recently brought a 6.6 KWh lithium-ion based rechargeable home battery back-up system to consumer market
that promises storage solution for the entire household. Moreover, households with battery systems have the advantage of generating additional income by selling surplus stored energy during peak periods to neighbours without storage facilities \cite{6552325}. Along with pricing incentives, scheduling capabilities, renewable energy source integration, load leveling and cost minimization options, home battery systems are not only cost-effective for the households in the long run, but also increase the social welfare for the entire energy generation and distribution system.

Saving goods or money for future use is an inherent characteristic of \textit{Homo economicus} \cite{Ramsey}. In macroeconomic theory, intertemporal trade \cite{macro} is defined as the transaction of goods or money across time when an agent is faced with the option of consuming and/or saving in the present with the aim of using the savings in the future \cite{fisher}. In order to maximize the benefits from energy storage, the household consumption profile must adapt to the changing patterns of demand and market prices. Macroeconomic theory can be applied to households with storage devices to arrive at an optimal consumption profile. Optimality could either be defined as minimizing consumption costs or balancing/leveling consumption to produce uniform demand but not necessarily both. Cost minimization involves storing as much energy as possible during off-peak hours when demand and prices are lower and using the stored energy during peak hours when demand and prices are higher and hence the resulting consumption profile cannot be expected to be uniform. Under this scheme, the household is the sole beneficiary and there is no incentive for the utility company to support this scheme as the resulting consumption profile, if not worse, is as non-uniform as the original energy requirements of the household. Consumption balancing/leveling on the other hand involves making the consumption profile as uniform as possible without incurring any additional costs. Under this scheme, there is an incentive for both the household and the utility company to support this scheme (reduction in consumption costs for the former and uniform demand response or balanced overall load for the latter). In order to achieve an optimal consumption profile irrespective of the type of optimality considered, households need a stategy that answers the key question of when and how much to charge or discharge their batteries, while also making sure that the battery is operated under proper conditions that extend its life span. Modeling energy consumption of households with storage devices as an intertemporal trading economy for consumption balancing/leveling leads to a class of optimization problems known as geometric programs (GPs) \cite{Boyd}, while cost minimization is solved using conventional linear programming subject to respective budget, consumption, storage and savings constraints.

Using micro/macro economic and game-theoretic concepts to study and model the dynamics of smart grid systems is a fairly recent approach. A market clearing auctioning approach towards buying and selling demand response as a public good has been studied in \cite{5978808}. A review of the impact of vehicle to grid technologies on distribution systems and utility interfaces is provided in \cite{6353961}. Deployment of optimal and autonomous incentive based ECS algorithm for smart grids without energy storage devices is discussed in \cite{5434752}. A comprehensive tutorial for applying game-theoretic methods to smart grid systems with respect to microgrid study, DSM and smart grid communications can be found in \cite{tutorial}. A non-cooperative game-theoretic approach to modeling DSM with energy storage devices for a whole loaclity is discussed in \cite{eps268360}. The paper studies the effect of multiple households with battery systems in the same neighborhood simultaneously opting for cost minimization scheme that could lead to extremely non-uniform demand resulting in grid failure and suggests a game-theoretic and machine learning based approach to arrive at a Nash equilibrium consumption point. The main contributions of this paper are as follows.
\begin{itemize}

\item A novel framework for modeling the energy consumption of households connected to the smart grid with energy storage systems as an intertemporal trading economy is proposed.

\item Optimal energy consumption of a household is defined in two different ways. In one scheme, the consumption of the household is balanced/leveled such that the utility company is presented with a demand that is as uniform as possible. In another scheme, the household consumption costs are minimized by storing energy during off-peak periods when demand and prices are lower and using the stored energy during peak periods when demand and prices are higher.

\item The preferences of the households when faced with a choice between consuming in the present to fulfill its current energy requirements, storing energy for future use and spending energy stored in the past are represented by appropriate utility functions using consumer theory.

\item Two different utility functions are introduced, one for the cost minimization scheme and another for the consumption balancing/leveling scheme that are optimized subject to respective budget, consumption, storage and savings constraints.

\item Consumption balancing/leveling is formulated as a geometric programming optimization problem while cost minimization is formulated as a linear programming optimization problem

\item For a given set of hourly day-ahead market energy prices set by the utility company, hourly energy requirements of the household and operational parameters of the battery system, the proposed model is able to achieve extremely low consumption PAR close to 1 under the consumption balancing/leveling scheme with reduction in consumption costs of about 8\% and presents the household with the least possible amount for electricity bill with reduction in consumption costs of about 12\% under the cost minimization scheme.

\end{itemize}

The rest of this paper is organized as follows. The system model is briefly described in Section II. The concept of intertemporal trade and its applicability to energy storage systems is analysed in detail in Section III. In Section IV, an introduction to consumer theory is provided and examples are given for graphically solving the optimal consumption profiles of two- and three-period models. In Section IV, consumption optimization is set up as linear progamming problem for cost minimization and geometric programming problem for load balancing. The simulation results are presented and analyzed in section V. Section VI concludes the paper.

\section{System Model}

Consider a smart grid system where a household is served by a utility company that exogenously provides energy. Additionally, households may also generate energy by means of privately-owned renewable energy sources such as solar panels, wind turbines, etc. Households amy be equipped with energy storage devices which may either be a dedicated battery system or a PHEV. Households may also have a smart meter installed with appliance scheduling capabilities that is connected to the power lines from energy sources.  Households have access to day-ahead hourly prediction prices issued by their utility companies so that they can schedule the use of their appliances accordingly and choose the most optimum strategy for charging and discharging their batteries. Each household also has accurate knowledge of its hourly energy requirements for the day. The energy requirements of the household may be the actual load generated by operation of appliances or it could also be the adjusted load after accounting for appliance scheduling and energy production from renewable sources.

We define a $N$-period model for the household as a 24 hour day that is split equally into $N$ intervals and each period is indexed by $\{1, 2, 3, \cdots, N\}$. Household defined time periods are synchronised with the periods set by the utility company for their dynamic pricing model. The price, energy requirement, consumption and state of battery storage in periods 1 through $N$ are denoted by $p_1, l_1, c_1, b_1$ through $p_N, l_N, c_N, b_N$. It must be noted that $b_i$ is not the amount of energy stored in the battery in period $i$, but is the charge level of the battery at the end of period $i$. The amount of energy stored in period $i$ is therefore given by the difference between charge levels at the end of period $i$ and the loss accounted charge levels at the end of period $i-1$, i.e., $b_i - b_{i-1}(1-r)$. Let $b_0$ denote the initial state or charge level of the battery before the beginning of the first period, $b_N$ the final state of the battery at the end of $N$ periods and $b_{max}$ its maximum charge levels or its capacity. Without any prior knowledge about the state of the battery before period 1 and after period $N$, we can set both $b_0$ and $b_N$ to zero. However, any arbitary value for $b_0$ and $b_N$ can be set in this model without loss of generality.

Let $r$ be the rate of storage loss per period in the battery that accounts for unavoidable self-discharge and other loss factors, meaning $E$ Wh of energy stored in one period is worth $E(1-r)$ Wh of energy in the next period and $E(1-r)^2$ after two periods and so on, or in other words, $(1-r)$ is the per period storage efficiency of the battery. Typical self discharge loss rates for lithium-based batteries are around 2\% to 3\% per month, while nickel-based batteries suffer higher losses at around 15\% to 30\% per month. Hence, by any conservative estimate, it quite safe to assume $r$ = 0.0001 for a 24-period model, while lower period models suffer from higher battery loss rates, for example, $r$ = 0.01 for a two-period model. Loss rates are also dependent upon the age of the battery, temperature fluctuations, state of the cycling period etc. We assume that the batteries have quick transfer rates, meaning they can be charged or discharged from one level to another within the duration of a time period. To further simplify the analysis, we also assume that charging and discharging are mutually exclusive within a time period and hence, batteries can only either be charged or discharged during a time period, but not both. In order to prolong longevity, every battery system must be operated within its state of charge, which is specified as a range of percentage of its capacity. For example, a Toyota Prius PHEV with 4.4 KWh lithium-ion battery pack, has state of charge between 40\% to 80\%. Battery systems can also be scheduled to charge and discharge at required times \cite{6531031} as programmable chargers for different battery types from various companies are available in the commercial market. 

Finally, we make a simplistic assumption that there are no externalities in the market, that is, each household cares only about the amount of energy that it consumes and is not concerned with the consumption of other households in the neighborhood, even though it may indirectly affect the market. We also assume that households are price takers, meaning that they take the prices in the market as fixed and act accordingly, and have no direct power (or at least believe that they have no power) to change the market prices.

\section{Intertemporal Trade}

Intertemporal trade is the transaction of goods across time in oder to benefit from the changing values of goods with time. In the context of optimal energy consumption with storage devices, during any time period, a household is faced with three consumption options. It can consume the exact amount of energy required for its household operations during that time period, or consume more than the amount of energy required, use the additional energy to charge its batteries and store it for future use or consume less than the amount of energy required and use the energy stored in the past by discharging the batteries. The consumption preferences of a household will therefore depend upon its own energy requirements at different time periods and storage loss rate. Given a energy requirement profile of the household along with its battery loss rate, intertemporal trade provides the bounds on consumption during every time period, also known as the budget constraint. We start with a simple two-period model to explain the concepts of intertemporal trade and generalize to a higher dimensional time period model.

\subsection{Two-period Model}

\begin{figure}[htpb]
	\centering
	\includegraphics[width=0.95\textwidth]{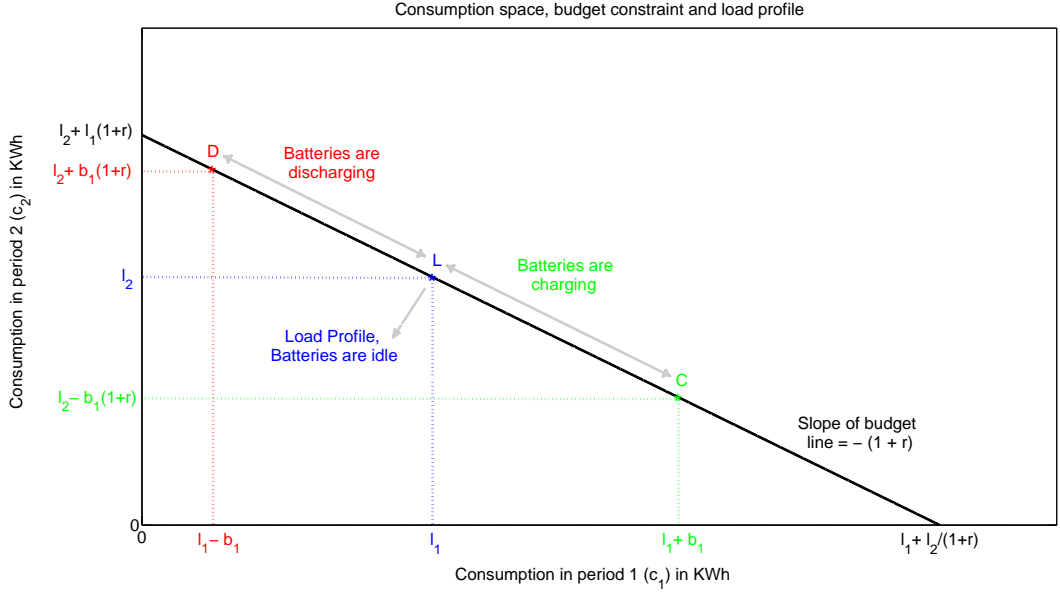}
	\caption{Consumption space, budget constraint and energy requirement profile of a household. The black line gives the budget constraint and its intercepts represent extreme points of consumption in different periods. Point L is the load or energy requirement profile when there is no storage and consumption is same as the load profile. Point C shows the level to which the batteries are charged in period 1, that is, the amount of increase in consumption during period 1 from $l_1$ to $l_1+b_1$,  is stored and used later to decrease the consumption in period 2 from $l_2$ to $l_2 - b_1(1-r)$. Similarly, point D shows the level to which the batteries are discharged.}
\label{fig1}
\end{figure}

Assume that households face only two time periods in a day, meaning the utility company sets prices only twice a day or for two periods, where period 1 occurs during off-peak hours when energy prices are low and period 2 occurs during peak hours when prices are high, and that the energy requirements and prices within these time periods are constant. More variations and finer resolution in pricing and energy requirements can be accommodated into the model by increasing the model order such that the energy requirements and prices are constant within those periods. This pricing mechanism provides incentive for the households to schedule consumption or store energy during off-peak periods. In period 1, the household consumes an amount equal to its energy requirements in that period with the option of spending any stored energy from the previous period and since the prices are lower than in next period, the household additionally also chooses to store energy by charging its batteries. Since there is no prior knowledge about stored energy before period 1, we set $b_0$ to zero, but any arbitary value will also work with this model and hence, this is not a restrictive asumption. The equation for consumption in period 1 is given by,
\begin{equation}
c_1 = l_1 + b_1 - b_0 (1-r) = l_1 + b_1.
\label{1}
\end{equation}
In period 2, the household can use the stored energy from period 1 to partially or fully meet its load by discharging the batteries. If there is no prior knowledge about the energy prices for the next day, $b_2$ can be set to zero at the end of period 2, but any arbitary value will also work without affecting the model. Thus, the equation for consumption in period 2 is given by, 
\begin{equation}
c_2 = l_2 + b_2 - b_1 (1-r) = l_2 - b_1 (1-r).
\label{2}
\end{equation}
Solving for $b_1$ from Eq. \eqref{2}, and substituting in Eq. \eqref{1}, we get,
\begin{equation}
c_1 = l_1 + \frac{l_2 - c_2}{(1-r)}.
\label{3}
\end{equation}
Rearranging the load and consumption terms in Eq. \eqref{3}, we arrive at the budget constraint of the household as shown in Eq. \eqref{4}.
\begin{equation}
c_1 + \frac{c_2}{(1 - r)} = l_1 + \frac{l_2}{(1 - r)}.
\label{4}
\end{equation}
The budget constraint of the household gives the present value (i.e., w.r.t period 1) of total consumption in terms of its present value of total energy requirements. This is illustrated graphically in Fig. \ref{fig1}.
Let $(c_1,c_2) \in \mathbb{R}_+^2$ be the consumption space. The budget constaint is a line in the consumption space whose intercepts are determined by the energy requirement profile of the household. The horizontal intercept $l_1 + l_2/(1-r)$ gives the amount of energy that is required in period 1 if there would be no consumption in period 2, that is, in addition to meeting its energy requirements during period 1, the household charges its batteries such that the energy requirement during period 2 is met using only the stored energy. Similarly, the vertical intercept $l_2 + l_1 (1-r)$ gives the amount of energy that is required in period 2 if there would be no consumption in period 1. The slope of the budget line is given by $-(1-r)$, or the negative of storage efficiency. The household can operate at any consumption profile ($c_1,c_2$), a point that is on the budget line (efficient) or in the region below the line (inefficient), but not above the line (unattainable). The point $L$ on the budget line is the household energy requirements or in other words, its load profile ($l_1,l_2$). The energy requirements of the household may be the actual load generated by operation of appliances or it could also be the adjusted load after accounting for appliance scheduling and energy production from renewable sources. When the household is operating at point L, its consumption is equal to its energy requirements during that period and the batteries are idle and not used. If period 1 occurs during off-peak hours when market prices are low, the household operates at point $C$ on the budget line, where, in addition to the normal energy requirements $l_1$, the consumption is $l_1 + b_1$ and the batteries are charged to $b_1$ to be used in period 2 when the market prices are higher. Similarly, during peak hours when market prices are high, the household operates at point $D$ where the energy stored in the batteries are used to reduce the consumption and costs. 

\subsection{N-period Model}

We extend the two-period model to a three-period model. The consumption equation for period 3 is given by,
\begin{equation}
c_3 = l_3 + b_3 - b_2(1-r) = l_3 - b_2(1-r).
\label{5}
\end{equation}
Solving for $b_2$ from Eq. \eqref{5}, and substituting in the consumption equation for period 2 given by Eq. \eqref{2}, we get,
\begin{equation}
c_2 = l_2 + b_2 - b_1(1-r) = l_2 + \frac{l_3 - c_3}{(1-r)} - b_1(1-r)
\label{6}
\end{equation}
Rearranging Eq. \eqref{6} and solving for $b_1$, we get,
\begin{equation}
b_1 = \frac{l_3}{(1-r)^2} + \frac{l_2}{(1-r)} - \frac{c_3}{(1-r)^2} - \frac{c_2}{(1-r)}.
\label{7}
\end{equation}
Substituting for $b_1$ in the consumption equation for period 1 given by Eq. \eqref{1}, we get,
\begin{equation}
c_1 = l_1 + b_1 - b_0(1-r) = l_1 + \frac{l_3}{(1-r)^2} + \frac{l_2}{(1-r)} - \frac{c_3}{(1-r)^2} - \frac{c_2}{(1-r)}.
\label{8}
\end{equation}
Rearranging the energy requirement and consumption terms in Eq. \eqref{8}, we arrive at the budget constraint of the household as shown in Eq. \eqref{9}.
\begin{equation}
c_1 + \frac{c_2}{(1-r)} + \frac{c_3}{(1-r)^2} = l_1 + \frac{l_2}{(1-r)} + \frac{l_3}{(1-r)^2} .
\label{9}
\end{equation}
Extending recursively to a N-period model, we can derive the general budget constraint hyperplane in an $N$-dimensional space as shown in Eq. \eqref{10}.
\begin{equation}
c_1 + \frac{c_2}{(1-r)} + \frac{c_3}{(1-r)^2} + \cdots + \frac{c_N}{(1-r)^{N-1}} = l_1 + \frac{l_2}{(1-r)} + \frac{l_3}{(1-r)^2} + \cdots + \frac{l_N}{(1-r)^{N-1}}.
\label{10}
\end{equation}
The problem faced by households can now be stated as follows: Given $N$ periods $\{1, 2, \cdots, N\}$, market prices $\mathbf{p} = [p_1, p_2, \cdots, p_N]^T$, energy requirements $\mathbf{l} = [l_1, l_2, \cdots, l_N]^T$ and battery loss rate $r$ per period, at which point on the budget hyperplane should the household operate, that is, how to choose an optimal consumption profile $\mathbf{c^*} = [c_1^*,c_2^*, \cdots, c_N^*]^T$, or in other words, when should the household charge or discharge its batteries and by how much ? The answer is given by consumer theory.

\section{Consumer Theory}
\begin{figure}[htpb]
	\centering
	\includegraphics[width=0.95\textwidth]{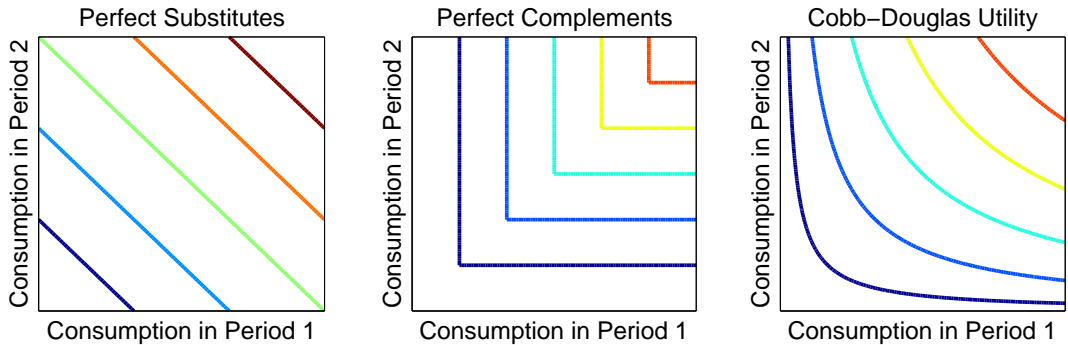}
	\caption{Examples of common utility functions such as perfect substitutes, perfect complements and Cobb-Douglas utility function with $\alpha_1,\alpha_2 = 0.5$. The indifference curves link points of equal utility in terms of consumption in periods 1 and 2.}
	\label{fig2}
\end{figure}

\begin{figure}[htpb]
	\centering
	\includegraphics[width=0.95\textwidth]{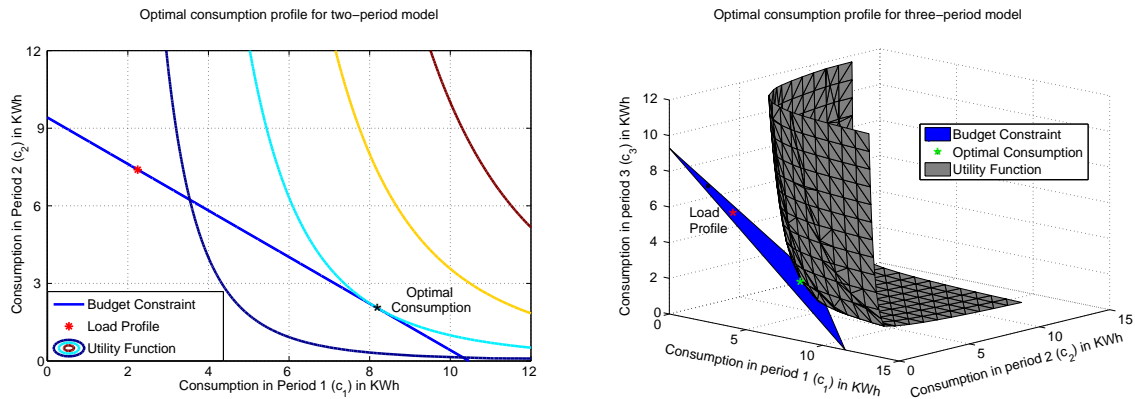}
	\caption{Graphical representation of optimal consumption point for two-period and three-period models. The optimal consumption point occurs when the budget hyperplane is tangential to the indifference surfaces of the utility function.}
	\label{fig3}
\end{figure}

Households make rational choices using consumption preferences which are defined in the consumption space. A preference relation on the consumption space specifies how a particular amount of consumption in one period is valued with respect to an amount of consumption in another time period. Formulating appropriate utility functions that reflect the preference relations of households with regards to their consumption over different time periods is of vital importance in modeling intertemporal trade. Some common utility functions for a two-period model are given below.
\begin{align}
u(c_1,c_2) &= c_1 + c_2, &\text{  (Perfect Substitutes)} \label{11} \\
u(c_1,c_2) &= min(c_1,c_2), &\text{  (Perfect Complements)} \label{12} \\
u(c_1,c_2) &= {c_1}^{\alpha_1} \times {c_2}^{\alpha_2} &\text{  (Cobb-Douglas Utility)} \label{13} \\
u(c_1,c_2) &= -(w_1c_1 + w_2c_2). &\text{  (Weighted Minimization)} \label{14}
\end{align}
Utility $u(c_1,c_2)$ as a function of consumption is usually visualized as isoquants or contours in the two dimensional consumption space with each consumption period on each of the axes and contour lines linking points of equal utility. The constant utility contour lines are known as indifference curves as they link points of equal preference, in other words, linking consumption points that are indifferent. Some common utility functions for a two-period model are shown in Fig. \ref{fig2}.
If a household does not care about consumption in individual periods, but is concerned only about the total consumption, the perfect substitute utility function in Eq. \eqref{11} captures this preference aptly as consumption in period 1 can be substituted for consumption in period 2. If the utility company employs a static pricing model, then households will tend to use this kind of utility function to optimize their consumption. If a household values consumption in period 1 with a certain minimum constraint on consumption in period 2, this preference is reflected in the perfect complement utility function in Eq. \eqref{12} If the household has renewable energy sources with a fixed amount of energy production, then the perfect complement utility function would be ideal for modeling this scenario. However, if a household values a certain share of consumption in period 1 ($\alpha_1$) in relation to consumption in period 2 ($\alpha_2$), the Cobb-Douglas utility function in Eq. \eqref{13} is best suited for modeling this preference. This kind of a utility function is applicable to households that try to balance/level their consumption to help the utility company by supplying a uniform demand. If a household does not care about consumption in individual periods, but is concerned only about minimizing its total consumption costs, then the weighted minimization utility function in Eq. \eqref{14} can be used with prices as weights. Since these utility functions capture the best possible trade-off between consuming and storing energy under different scenarios while taking into account the current and future load and market prices, the optimal consumption point of a household is achieved when its utility function is maximized subject to its budget constraint. For a two-period model, the optimal consumption point occurs when the budget line is tangential to the highest indifference curve of the utility function and for a three-period model, the optimal consumption point occurs when the budget plane is tangential to the highest indifference surface of the utility function. This is graphically demonstrated using a Cobb-Douglas utility function in Fig. \ref{fig3}.
Once the nature of the optimality is decided, we can arrive at the optimal consumption point for a household by maximizing its utility subject to the budget constraint. In the following section, we formulate the optimization problem for minimizing the consumption costs of a household and for balancing consumption over a period of time such that the household presents the utility company with a uniform demand.

\section{Consumption Optimization}

The objective of optimal consumption can be of two types. Equipped with a storage device, the goal of the household could either be to minimize its consumption costs or to balance/level its consumption. Cost minimization involves storing as much energy as possible during off-peak hours when demand and prices are lower and using the stored energy during peak hours when demand and prices are higher and hence the resulting consumption profile cannot be expected to be uniform. If a household were to possess a battery system with enormous capacity, during period $i$ when the price is lowest, it could consume and store the loss accounted equivalent of its entire energy requirements for the next $N-i$ periods. Therefore, under this scheme, the household is the sole beneficiary and there is no incentive for the utility company to support this scheme as the resulting consumption profile, if not worse, is at least as non-uniform as the original energy requirements of the household. Consumption balancing/leveling on the other hand involves making the consumption profile as uniform as possible without incurring any additional costs and hence, there is an incentive for both the household and the utility company to support this scheme (reduction in consumption costs for the former and uniform demand for the latter). Thus, we see that the two kinds of optimality are non-compatible with each other and hence it would be logical to solve for these optimization problems separately. However, a household could try to balance/level the consumption as much as possible while simultaneously trying to keep the consumption costs as low as possible, but this is beyond the scope of this paper and there is no guarantee for the existence of a unique optimum that will jointly optimize both the objectives.

\subsection{Cost Minimization}

In this scheme, the household aims to minimize its consumption costs. The weighted minimization utility function can be used for this purpose with day ahead market energy prices as weights. Since the objective function is a linearly weighted function of consumption, this optimization can be formulated as a linear programming problem. The optimization problem for cost minimization over $N$ time periods is given by,
\begin{equation}
\underset{\mathbf{c}}{\textbf{Max:    }} -(p_1c_1 + p_2c_2 + \cdots + p_nc_N) = -\mathbf{p}^T\mathbf{c}.
\label{15}
\end{equation}
subject to the budget constraint given in Eq. \eqref{10}, consumption constraints and storage constraints.

\subsubsection{Consumption Constraints}

Consumption constraints arise due to limits on the nature of consumption. For example, consumption must be non-negative during all time periods, i.e., $c_i \geq 0$. Additionally, at any time period, consumption has to be greater than the energy requirements minus the maximum amount of stored energy that can be carried over from the previous periods, i.e., $l_i - b_{max}(1+r) \leq c_i$. On the other hand, consumption cannot exceed the sum of energy requirements and battery capacity, i.e., $c_i \leq l_i + b_{max}$. Thus, the consumption constraints for each time period can be expressed in vector notation as shown below.
\begin{equation}
\begin{bmatrix}
\text{max}(l_1-b_{max}(1-r), 0) \\
\text{max}(l_2-b_{max}(1-r), 0) \\
\vdots \\
\text{max}(l_N-b_{max}(1-r), 0)
\end{bmatrix}
\leq
\begin{bmatrix}
c_1 \\
c_2 \\
\vdots \\
c_N
\end{bmatrix}
\leq
\begin{bmatrix}
l_1+b_{max} \\
l_2+b_{max} \\
\vdots \\
l_N+b_{max}
\end{bmatrix}
\Longrightarrow \mathbf{lb} \leq \mathbf{c} \leq \mathbf{ub}.
\label{16}
\end{equation}
Thus, the consumption profile is restricted and is bounded from below by $\mathbf{lb}$ and from above by $\mathbf{ub}$.

\subsubsection{Storage Constraints}

Storage constraints arise due to natural limits on operation of battery systems. To simplify the analysis, we can assume that state of charge of battery systems used here is between 0\% to 100\% without loss of generality. Therefore, at the end of each time period, the amount of energy stored in the battery must be within its storage limits, i.e., $0 \leq b_i \leq b_{max}$. Actual state of charge values can be incorporated into this model by appropriately scaling the lower and upper limits. If the household uses its PHEV for energy storage purposes, there will be additional constraints on the battery usage such as specific time periods when the batteries can be used, the minimum amount of charge levels needed during certain time periods etc. Such constraints can also be incorporated into this model as long as the required charge levels in the battery can be specified in bounded form for each time period.

For period 1, $0 \leq b_1 \leq b_{max}$. From the consumption equation developed in section III, $b_1 = c_1 - l_1$ and therefore, $l_1 \leq c_1 \leq l_1 + b_{max}$. Similarly, for period 2, $0 \leq b_2 \leq b_{max}$ and hence, $0 \leq c_2 - l_2 + b_1(1-r) \leq b_{max}$. Substituting for $b_1$, we have $l_1(1-r) + l_2 \leq c_1(1-r) + c_2 \leq l_1(1-r) + l_2 + b_{max}$. Extending to N periods, we have,
\begin{align}
l_1(1-r)^{N-1} + l_2(1-r)^{N-2} + \cdots + l_N &\leq c_1(1-r)^{N-1} + c_2(1-r)^{N-2} + \cdots + c_N, \label{17} \\
c_1(1-r)^{N-1} + c_2(1-r)^{N-2} + \cdots + c_N &\leq l_1(1-r)^{N-1} + l_2(1-r)^{N-2} + \cdots + l_N + b_{max}. \label{18}
\end{align}

The lower limits for storage constraints during each period can therefore be represented as follows.
\begin{equation}
\begin{bmatrix}
l_1\\
l_1(1-r)+l_2 \\
\vdots \\
l_1(1-r)^{N-1}+\cdots+l_N
\end{bmatrix}
\leq
\begin{bmatrix}
1 & 0 & \cdots & 0 \\
(1-r) & 1 & \cdots & 0 \\
\vdots & \vdots & \ddots & \vdots \\
(1-r)^{N-1} & (1-r)^{N-2} & \cdots & 1
\end{bmatrix}
\begin{bmatrix}
c_1 \\
c_2 \\
\vdots \\
c_N
\end{bmatrix}
\Longrightarrow \mathbf{a} \leq \mathbf{Rc}.
\label{19}
\end{equation}

The upper limits for storage constraints during each period can be represented as follows.
\begin{equation}
\begin{bmatrix}
1 & 0 & \cdots & 0 \\
(1-r) & 1 & \cdots & 0 \\
\vdots & \vdots & \ddots & \vdots \\
(1-r)^{N-1} & (1-r)^{N-2} & \cdots & 1
\end{bmatrix}
\begin{bmatrix}
c_1 \\
c_2 \\
\vdots \\
c_N
\end{bmatrix}
\leq
\begin{bmatrix}
l_1+b_{max}\\
l_1(1-r)+l_2+b_{max} \\
\vdots \\
l_1(1-r)^{N-1}+\cdots+l_N+b_{max}
\end{bmatrix}
\Longrightarrow \mathbf{Rc} \leq \mathbf{d}.
\label{20}
\end{equation}
Thus, the optimization problem can be formally stated as follows :
\begin{align}
\text{Mazimize} & \ \ \ -(p_1c_1 + p_2c_2 + \cdots + p_Nc_N) = -\mathbf{p}^T\mathbf{c} \notag \\
\text{subject to} & \ \ \ c_1 + \frac{c_2}{(1-r)} + \cdots + \frac{c_N}{(1-r)^{N-1}} = l_1 + \frac{l_2}{(1-r)} + \cdots + \frac{l_N}{(1-r)^{N-1}}, \notag \\
& \ \ \ \mathbf{lb} \leq \mathbf{c} \leq \mathbf{ub}, \ \ \ \ \ \mathbf{a} \leq \mathbf{Rc} \leq \mathbf{d}. \label{21}
\end{align}

\subsection{Consumption Balancing/Leveling}

In this scheme, the household aims to balance/level its consumption such that the utility company is presented with a uniform demand. The Cobb-Douglas utility function is apt for representing how households value a certain share of consumption in every period depending upon the energy requirements and prices in order to even out overall consumption. The Cobb-Douglas utility is also posynomial function and when used as a cost function, leads to a special class of optimization problems known as geomteric programs \cite{Boyd}. The parameter $\alpha_i$ in the Cobb-Douglas utility function for period $i$ is chosen such that it represents the normalized cost of consumption in all time periods excluding $i$ and by constraining $\alpha_1 + \alpha_2 + \cdots + \alpha_n = 1$, the peaks in consumption are flattened. For example, in a two period model, $\alpha_1 = p_2l_2/(p_1l_1 + p_2l_2)$ and $\alpha_2 = p_1l_1/(p_1l_1 + p_2l_2)$. Since the objective function is a posynomial function \cite{Boyd} of consumption, this optimization can be formulated a geometric programming problem. The optimization problem for balancing/leveling consumption over $N$ time periods is given by :
\begin{align}
\underset{\mathbf{c}}{\textbf{Max:     }} u(c_1,c_2,\cdots,c_n) &= c_1^{\alpha_1} \times c_2^{\alpha_2} \times \cdots \times c_n^{\alpha_n} = \prod_{i=1}^{n} c_i^{\alpha_i}, \text{    where} \notag \\
\alpha_i &= \frac{\sum_{j=1,j \neq i}^{n} p_j l_j}{(n-1)\sum_{i=1}^{n} p_i l_i}. \label{22}
\end{align}
subject to the budget constraint given in Eq. \eqref{10}, consumption constraints, storage constraints and savings constraint. Since consumption balancing/leveling is formulated as a geometric programming problem, we have to convert the consumption and storage constraints as posynomial inequalities even though the constraints remain the same as in cost minimization problem.

\subsubsection{Consumption Constraints}

We convert the linear consumption constraints to posynomial inequalities. Thus, $\text{max}(l_1-b_{max}(1-r), 0) \leq c_1$ becomes, $c_1^{-1}\text{max}(l_2-b_{max}(1+r), \epsilon) \leq 1$ with $\epsilon \rightarrow 0$. Similarly, the upper limit $c_1 \leq (l_1+b_{max})$ becomes $c_1(l_1+b_{max})^{-1} \leq 1$. Thus the posynomial inequality consumption constraints can be represented as follows.
\begin{equation}
\begin{bmatrix}
c_1^{-1}\text{max}(l_2-b_{max}(1+r), \epsilon) \\
c_2^{-1}\text{max}(l_2-b_{max}(1-r), \epsilon) \\
\vdots \\
c_N^{-1}\text{max}(l_N-b_{max}(1-r), \epsilon)
\end{bmatrix}
\leq
\begin{bmatrix}
1 \\
1 \\
\vdots \\
1
\end{bmatrix}
\Longrightarrow \mathbf{cl_c} \leq \mathbf{1}
,
\begin{bmatrix}
c_1(l_1+b_{max})^{-1} \\
c_2(l_2+b_{max})^{-1}\\
\vdots \\
c_N(l_N+b_{max})^{-1}
\end{bmatrix}
\leq
\begin{bmatrix}
1 \\
1 \\
\vdots \\
1
\end{bmatrix}
\Longrightarrow \mathbf{cu_c} \leq \mathbf{1}.
\label{23}
\end{equation}

\subsubsection{Storage Constraints}

The lower limits for storage constraints as derived for the cost minimization scheme cannot be directly converted to posynomial inequality form and hence, we modify the constraints in terms of battery capacity to fit the posynomial inequality form. Thus, Eq. \eqref{19} can be modified as $c_n^{-1}\text{max}(l_n-b_{max}(1+r)-\cdots-b_{max}(1+r)^{n-1},\epsilon)$. The lower limits for storage constraints during each period can therefore be represented as follows.
\begin{equation}
\begin{bmatrix}
c_1^{-1}l_1 \\
c_2^{-1}\text{max}(l_2-b_{max}(1-r),\epsilon) \\
c_3^{-1}\text{max}(l_3-b_{max}(1-r)-b_{max}(1-r)^2,\epsilon) \\
\vdots \\
c_n^{-1}\text{max}(l_n-b_{max}(1-r)-\cdots-b_{max}(1-r)^{n-1},\epsilon)
\end{bmatrix}
\leq
\begin{bmatrix}
1 \\
1 \\
1 \\
\vdots \\
1
\end{bmatrix}
\Longrightarrow \mathbf{cl_s} \leq \mathbf{1}.
\label{24}
\end{equation}

The upper limits for storage constraints can be converted to posynomial inequalities directly. Thus, the constraint for period 2 is converted from $(c_1(1-r)+c_2 \leq l_1(1-r)+l_2+b_{max}$ to $(c_1(1-r)+c_2)(l_1(1-r)+l_2+b_{max})^{-1} \leq 1$. The upper limits for storage constraints during each period can be represented as follows.
\begin{equation}
\begin{bmatrix}
c_1(l_1+b_{max})^{-1} \\
(c_1(1-r)+c_2)(l_1(1-r)+l_2+b_{max})^{-1} \\
\vdots \\
(c_1(1-r)^{n-1}+\cdots+c_n)(l_1(1-r)^{n-1}+\cdots+l_n+b_{max})^{-1} \\
\end{bmatrix}
\leq
\begin{bmatrix}
1 \\
1 \\
\vdots \\
1
\end{bmatrix}
\Longrightarrow \mathbf{cu_s} \leq \mathbf{1}.
\label{25}
\end{equation}

\subsubsection{Savings Constraints}

In the consumption balancing/leveling scheme, we can also add a savings constraint that restricts the balanced/leveled consumption profile such that the household incurs no additional cost for balancing/leveling the consumption . Without this constraint, there will be no incentive for the household to balance/level its consumption. In order for total cost savings to be non-negative, we have, 
\begin{equation}
\mathbf{p}^T\mathbf{c} \leq \mathbf{p}^T\mathbf{l} \Longrightarrow \mathbf{p}^T\mathbf{c}/\mathbf{p}^T\mathbf{l} \leq \mathbf{1}.
\label{26}
\end{equation}
Thus, the optimization problem can be formally stated as follows :
\begin{align}
\text{Mazimize} & \ \ \ u(c_1,c_2,\cdots,c_n) = c_1^{\alpha_1} \times c_2^{\alpha_2} \times \cdots \times c_n^{\alpha_n} = \prod_{i=1}^{n} c_i^{\alpha_i}, \text{    where} \notag \\
& \ \ \ \alpha_i = \frac{\sum_{j=1,j \neq i}^{n} p_j l_j}{(n-1)\sum_{i=1}^{n} p_i l_i}. \notag \\
\text{subject to} & \ \ \ c_1 + \frac{c_2}{(1-r)} + \cdots + \frac{c_N}{(1-r)^{N-1}} = l_1 + \frac{l_2}{(1-r)} + \cdots + \frac{l_N}{(1-r)^{N-1}}, \notag \\
& \ \ \ \mathbf{cl_c} \leq \mathbf{1}, \ \ \ \mathbf{cu_c} \leq \mathbf{1}, \ \ \ \mathbf{cl_s} \leq \mathbf{1}, \ \ \ \mathbf{cu_s} \leq \mathbf{1},  \ \ \ \mathbf{p}^T\mathbf{c}/\mathbf{p}^T\mathbf{l} \leq \mathbf{1}. \label{27}
\end{align}

\subsection{Nature of Optimization Problems}

Cost minimization is formulated as a linear programming problem which belongs to the class of convex optimization problems. The linear cost function is convex and the linear constrains create a feasible region that is a convex polyhedron. As long as the constraints are not mutually inconsistent, the global optimum solution always exists and every local minima is also a global minima. Moreover, for strictly convex cost functions, if a minimum exists, then that minimum is also unique. Effective computational methods exist that can solve linear programming problems in polynomial time.

Consumption balancing/leveling is formulated as a geometric programming problem due to the posynomial form of the Cobb-Douglas utility function. Geometric programs are not convex, but can be converted into a convex programming problem by applying logarithmic transformation. Standard interior-point algorithms can solve geometric programs with 1000 variables and 10000 constraints in less than a minute \cite{Boyd}. Interior-point methods for geometric programs are quick, reliable, robust, efficient, require no initial guess or starting point and have been proven to have polynomial time complexity. As long as the problem is not infeasible (caused by mutually inconsistent constraints), global optimum exists and can always be found.

\section{Simulation Results}

Let us assume that the utility company charges households with energy prices based on the USA New England hourly real-time prices of 1st January, 2011 \cite{USA_NE}. We model the daily energy requirements of households with usage-statistics-based load model proposed in \cite{Richardson}. This model simulates daily energy requirements with one hour time resolution through simulation of appliance use and also by taking into account simulated resident activity in households.  It must be noted that these are merely representative values of market prices and energy requirements and in principle, any other set of prices and energy requirement values can be used to study consumption optimization. We solve the optimization problems formulated in Eq. \eqref{21} and Eq. \eqref{26} from the previous section using CVX \cite{cvx}, a MATLAB package for specifying and solving convex programs \cite{gb08}. Minimizing the cost of consumption is formulated as a linear programming problem and is therefore a stright forward case of convex programming. Consumption balancing/leveling on the other hand is formulated as a geometric programming problem which is not convex, but can be converted into a convex programming problem by applying logarithmic transformation. However, CVX allows geometric programs to be constructed in their native non-convex form, transforms them automatically to a solvable convex form, and translates the numerical results back to the original problem.

\subsection{24-period Model}

\begin{figure}[htpb]
\centering
	\includegraphics[width=0.99\textwidth]{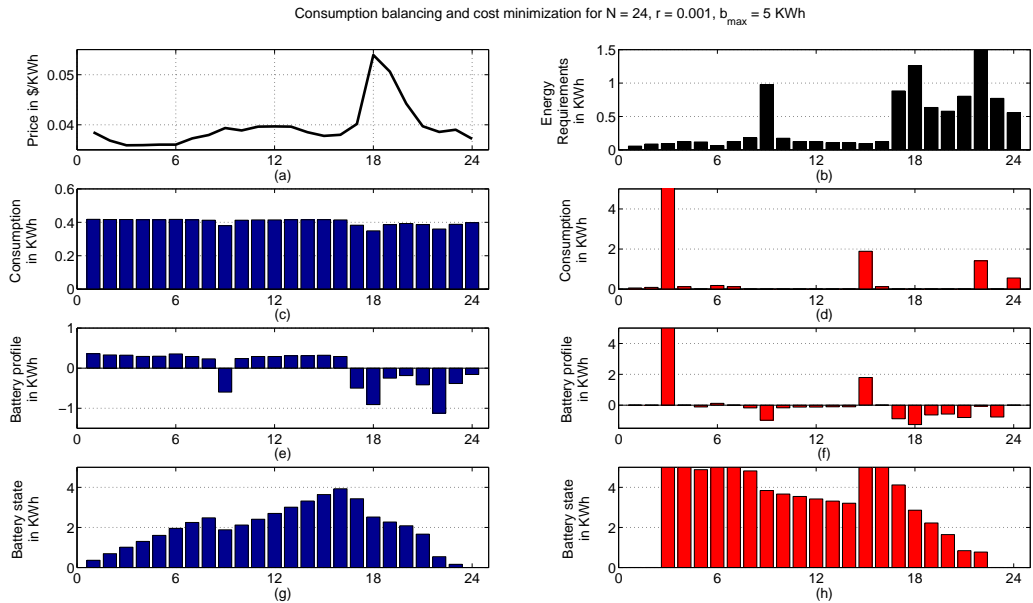}
	\caption{Consumption balancing/leveling and cost minimization for a 24-period model with $r = 0.001$ and $b_{max} = 5$ KWh. The day-ahead hourly market energy prices (a) and hourly energy requirements (b) are shown in black. The results for consumption balancing/leveling, (c), (e) and (g), are shown in blue while the results for cost minimization, (d), (f) and (g) are shown in red. Under the consumption balancing/leveling scheme, the household consumption is very uniform (PAR = 1.0390) over all time periods with reduction in consumption costs of about 5.5\% and the batteries are used at all time periods. Under the cost minimization scheme, reduction in consumption costs is around 11.5\% and the batteries are charged only during periods when prices are low and are discharged when prices are high.}
	\label{fig4}
\end{figure}

Given the day-ahead hourly market energy prices and hourly household energy requirements, we separately solve the 24-dimensional optimization problems for cost minimization and consumption balancing/leveling with battery loss rate of $r = 0.001$ and battery capacity of $b_{max} = 5$ KWh. The results are presented in Fig. \ref{fig4}. 
From Fig. \ref{fig4}, under the consumption balancing/leveling scheme, we see that the consumption of the household is very uniform over time (c) with PAR = 1.0390, even though the energy requirement profile (b) is highly non-uniform. The battery profile (e) shows the levels to which the batteries are charged or discharged during each time period and the battery state (g) shows the current energy levels of the battery at the end of each time period. We see that in the consumption balancing/leveling scheme, the batteries are used continuously during all time periods in order to keep the consumption uniform. A reduction of about 5.5\% in consumption costs is achieved. Under the cost minimization scheme a reduction of about 11.5\% in consumptions costs is achieved, which is the maximum possible reduction in costs for the given prices, energy requirements and battery parameters. We see that consumption (d) is less uniform when compared to the energy requirements (b). This is because the household consumes more to charge the batteries when the prices are low and consumes less when the batteries are discharging. Thus, we see from the battery profile (f) and state (h) that the batteries are used only during those time periods when the prices are advantageous for either charging or discharging in order to mazimize the gains obtained from storing energy. See Appendix for detailed analysis on the price conditions that determine when the batteries are charged and discharged for cost minimization.

\subsection{Effect of Battery Loss Rate}

The battery loss rate $r$, affects the optimal consumption point both for consumption balancing/leveling and cost minimization. The results for optimal consumption with battery loss rate of $r = 0.01$ and battery capacity of $b_{max} = 5$ KWh are presented in Fig. \ref{5}. 
\begin{figure}[htpb]
\centering
	\includegraphics[width=0.95\textwidth]{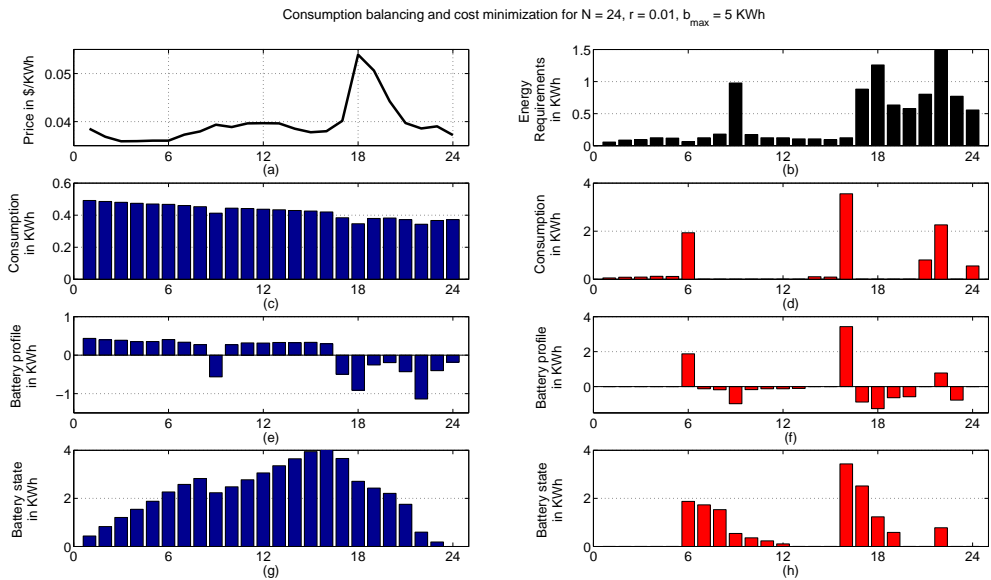}
	\caption{Consumption balancing/leveling and cost minimization for a 24-period model with $r = 0.01$ and $b_{max} = 5$ KWh. The day-ahead hourly market energy prices (a) and hourly energy requirements (b) are shown in black. The results for consumption balancing/leveling, (c) and (e), are shown in blue while the results for cost minimization, (d) and (f) are given in red. In the consumption balancing/leveling scheme, reduction in consumption costs is about 1\%, and the consumption is fairly uniform (PAR = 1.1597), but not quite as uniform when the battery loss rate is 0.001. In the cost minimization scheme, the reduction in consumption costs is around 8.5\% and we see that the batteries are used less frequently due to higher battery losses.}
	\label{fig5}
\end{figure}
We see that for the consumption balancing/leveling scheme, the household consumption (c) is fairly uniform with PAR = 1.1597, when compared to the energy requirements, but not as much when compared to the consumption with lower battery loss rate of $r = 0.001$ with PAR = 1.0390 as shown in Fig. \ref{4}(c). Since the loss rate is higher, the household has to consume more than the required amount to flatten the peaks in the energy requirements, thereby resulting in higher non-uniformity in overall consumption. Similarly, the reduction in consumption costs also drop from 5.5\% to about 1\% with increasing battery losses. Under the cost minimization scheme, we see that batteries are charged less frequently when compared to Fig. \ref{4}(g) and (h). This is because, the gains obtained from savings is nullified by the higher battery loss rate and hence the household prefers to charge and use the batteries only when the prices are very low to justify the usage of batteries with higher loss rates. In this case too, the reduction in consumption costs drop from 11.5\% to about 8.5\% with increasing battery loss rates.

The results for the effect of battery loss rate on the reduction in consumption costs obtained under cost minimization scheme for various time-period models at $b_{max} = 6.5$ KWh is shown in Fig. \ref{6}.
\begin{figure}[htpb]
\centering
	\includegraphics[width=0.95\textwidth]{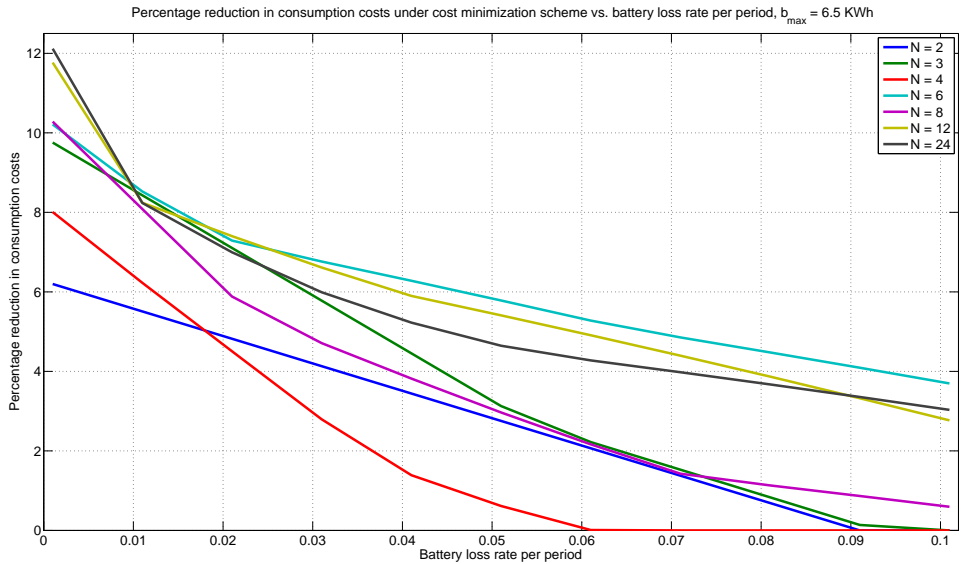}
	\caption{Effect of battery loss rate on reduction in consumption costs obtained under cost minimization scheme for various time-period models at $b_{max} = 6.5$ KWh. Savings decrease with increasing battery loss rates and tend to go towards zero quickly for lower period models.}
	\label{fig6}
\end{figure}
For example, to analyse a two-period model, we divide the USA NE hourly prices into two time periods with period 1 running from midnight until noon and period 2 from noon until midnight and set market prices $p_1$ and $p_2$ by averaging the prices over those time periods. Any arbitary time period division is possible depending upon the utility company's definition of peak and off-peak hours without loss of generality. The simulated energy requirements of the household is aggregated over the time periods and given by $l_1$ and $l_2$. Similarly, for a three-period model, we split the USA NE hourly prices into three time periods and set market prices $p_1$, $p_2$ and $p_3$ by averaging the prices over those time periods. The simulated energy requirements of the household is aggregated over time periods and given by $l_1$, $l_2$ and $l_3$. From Fig. \ref{6}, we see that the reduction in consumption costs go to zero quickly when the battery loss rate increases for lower period models such as $N = 2, 3$ and 4. For higher period models, the reduction in consumption costs monotonically decrease with increase in battery loss rate and tend towards zero for higher battery loss rates. This is expected because higher battery loss rates don't provide any incentive for the households to store energy and hence the reduction in consumption costs are lesser due to absence of storage. Moreover, reduction in consumption costs is not directly proportional to the model order because of the effects of averaging prices and aggregating energy requirements for lower order models and also because of the discontinuity introduced in prices and energy requirements due to time period division, but the general trend is that reduction in consumption costs decrease with increasing battery loss rates.

The results for the effect of battery loss rate on the nature of consumption under the consumption balancing/leveling scheme for various time-period models at $b_{max} = 5.5$ KWh is shown in Fig. \ref{7}. The extent of uniformity of the consumption under this scheme can be assessed using the peak to average consumption ratio and the variance of the consumption. 
\begin{figure}[htpb]
\centering
	\includegraphics[width=0.95\textwidth]{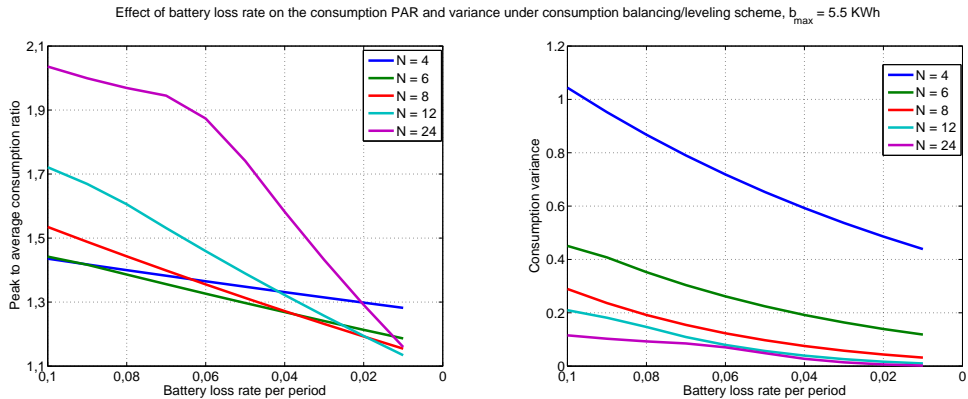}
	\caption{Effect of battery loss rate on the peak to average consumption ratio and the variance of consumption under the cost minimization scheme for various time-period models at $b_{max} = 5.5$ KWh. With decreasing battery loss rates, we see that the consumption PAR tends towards 1 and the consumption variance tends towards 0.}
	\label{fig7}
\end{figure}
The consumption PAR provides an estimate of the amount by which the maximum consumption is higher than the average consumption over a period of time. Along with the variance of the consumption, PAR is useful for analysing the uniformity of the consumption resulting from the consumption balancing/leveling scheme. From Fig. \ref{fig7}, we see that both the consumption PAR and variance of the consumption decrease monotonically with decreasing battery loss rates. Thus, lesser the battery loss rate, higher the uniformity of resulting optimal consumption.

\subsection{Effect of Battery Capacity}

The capacity of the battery $b_{max}$ also affects the optimal consumption point. The results for the effect of battery capacity on the reduction in consumption costs obtained under both the consumption balancing/leveling scheme and cost minimization scheme for various time-period models at $r = 0.01$ is shown in Fig. \ref{8}.
\begin{figure}[htpb]
\centering
	\includegraphics[width=0.49\textwidth]{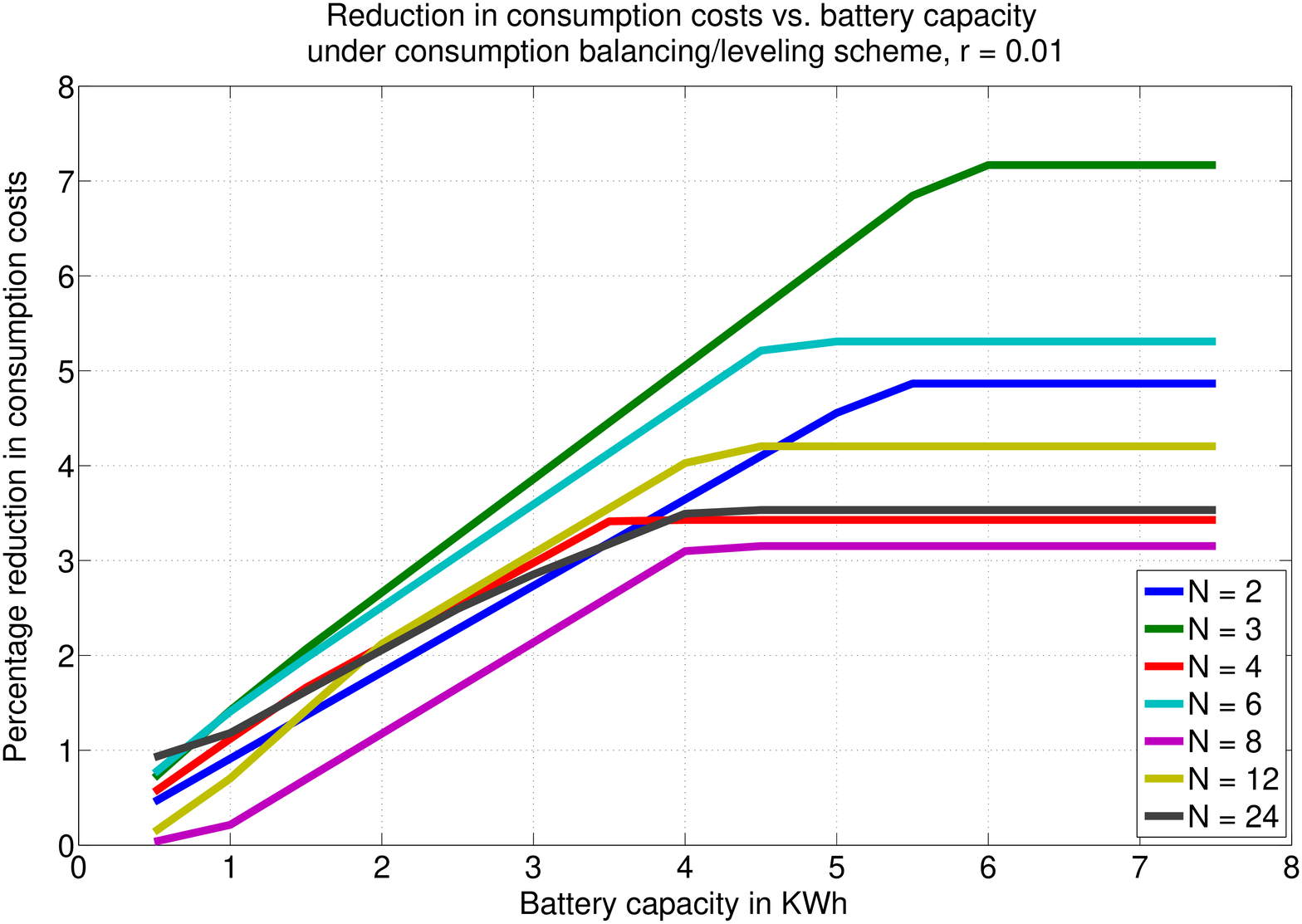}
	\includegraphics[width=0.49\textwidth]{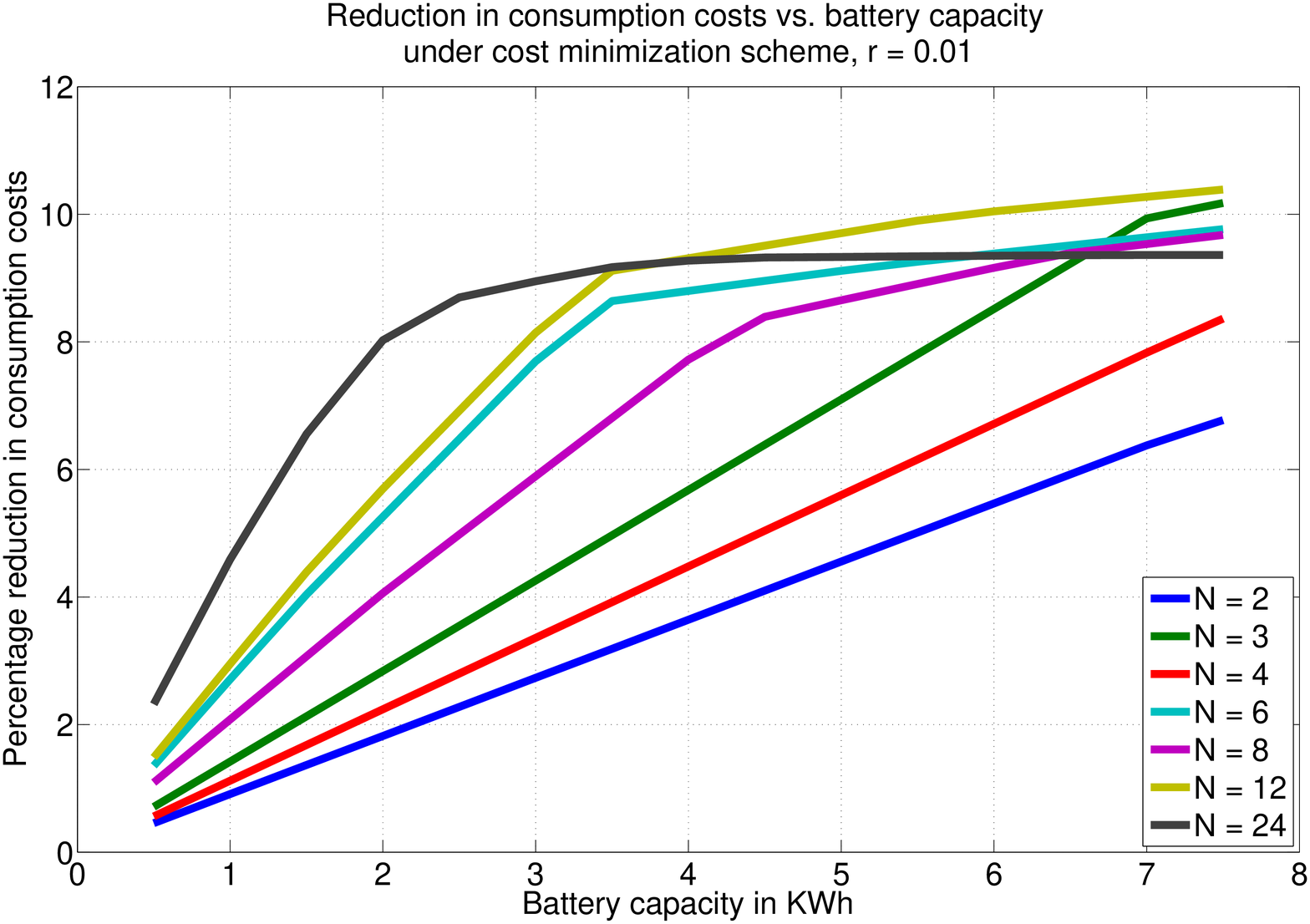}
	\caption{Effect of battery capacity on reduction in consumption costs under the consumption balancing/leveling (a) and cost minimization (b) scheme for various time-period models at $r = 0.01$. Reduction in consumption costs obtained under both schemes increase with battery capacity, but the reduction in consumption costs saturate quickly in consumption balancing/leveling scheme when compared to the cost minimization scheme.}
	\label{fig8}
\end{figure}
We see that the reduction in consumption costs obtained in both schemes monotonically increase with the battery capacity. This is expected because, higher the capacity of the battery, higher the amount of energy that can be stored and used later. In the consumption balancing/leveling scheme (a), we also see that the reduction in consumption costs saturate quickly, implying that beyond a certain battery capacity, no amount of increase in the capacity, gives a higher reduction in consumption costs as the consumption balancing/leveling constraint is the main focus of this scheme which cannot be compromised for reduction in consumption costs. In the cost minimization scheme (b), we see that higher period models tend to saturate reduction in consumption costs around a particular value of battery capacity while lower periods tend to saturate in the long run.

The results for the effect of battery capacity on the consumption PAR and consumption variance under the consumption balancing/leveling scheme for various time-period models at $r = 0.03$ is shown in Fig. \ref{9}. 
\begin{figure}[htpb]
\centering
	\includegraphics[width=0.95\textwidth]{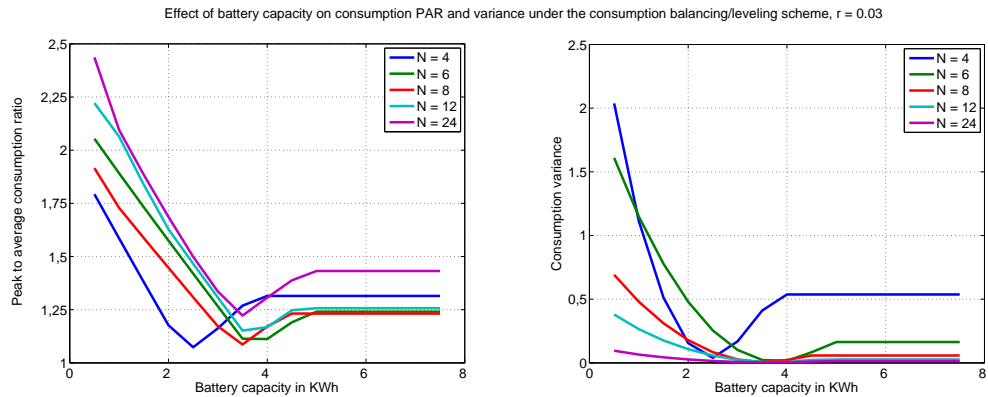}
	\caption{Effect of maximum battery capacity on the peak to average consumption ratio and the variance of consumption under the cost minimization scheme for various time-period models at $r = 0.03$. Non-uniformity of the consumption decreases with increasing battery capacity before increasing slightly and saturating.}
	\label{fig9}
\end{figure}
We see that both the consumption PAR and consumption variance tend to reach a minimum at around $b_{max} = 3$ KWh and increase slightly before saturating at higher battery capacities. This marginal increase in the non-uniformity of consumption can be attributed to the fact that higher capacity gives more flexibility for the household in terms of the maximum amount of energy that can be stored and hence the increase in the variance of consumption.

\subsection{Summary of Results}

For a given set of hourly market energy prices set by the utility company, hourly energy requirements of the household and battery parameters,

\begin{itemize}

\item Consumption balancing/leveling scheme achieves extremely low consumption PAR values close to 1, making the consumption almost perfectly uniform.

\item Cost minimization scheme achieves about 12\% reduction in consumption costs, whereas consumption balancing/leveling scheme achieves about 8\% with battery loss rate $r = 0.001$.

\item Reduction in consumption costs increases with decreasing battery loss rates and increasing battery capacities for both schemes.

\item Uniformity in consumption increases with decreasing battery loss rates and increasing battery capacities for consumption balancing/leveling scheme.

\item Batteries are charged and discharged more frequently in the consumption balancing/leveling scheme than in cost minimization scheme.

\end{itemize}

\section{Conclusion}

A novel framework for modeling energy consumption of households connected to the smart grid with energy storage devices as an intertemporal trading economy is proposed. The model is also applicable for households with renewable energy production sources and energy storage systems such as dedicated batteries or PHEVs. Due to the dynamic nature of market energy prices and demand, the household is faced with a choice between consuming in the present to fulfill its current energy requirements, storing energy for future use and spending energy stored in the past. The resulting consumption preferences of the household are modeled as utility functions using consumer theory. Two different utility functions for optimizing household energy consumption are introduced, where, one aims to minimize household consumption costs, while the other aims to balance/level the consumption such that the utility company is presented with a uniform demand. In the cost minimization scheme, the household is the sole beneficiary as it has to pay the least possible cost for energy consumption, but there is no incentive for the utility company as the resulting consumption is very non-uniform. On the other hand, in the consumption balancing/leveling scheme, the household gains from reduced consumption costs while the utility company is presented with a demand that is as uniform as possible and therefore this scheme is beneficial to both. Cost minimization is formulated as a linear programming problem and consumption balancing/leveling is formulated as a geometric programming problem. Both optimization problems are solved subject to respective budget, consumption, storage and savings constraints. Simulation results show that the proposed model achieves extremely low consumption PAR values close to 1 in the consumption balancing/leveling scheme with reduction in consumption costs of about 8\% and presents the household with the least possible amount for electricity bill with about 12\% reduction in consumption costs in the cost minimization scheme.

\appendix

The savings obtained from energy storage is dependent on the price variations between different time periods. In order to benefit from storing energy, total cost savings $S$ must be non-negative and therefore, the price weighted difference between the energy requirements and total consumption must be positive over $N = {1, 2, \cdots, N}$ time periods. Thus, the equation for cost savings is given by,
\begin{equation}
S = \mathbf{p}^T(\mathbf{l} - \mathbf{c}) = p_1(l_1 - c_1) + p_2(l_2 - c_2) + \cdots + p_N(l_N - c_N) \geq 0.
\label{28}
\end{equation}
From the consumption equation for any time period $i$ in section III, we have $l_i - c_i = b_{i-1}(1-r) - b_i$. Substituting this in Eq. \eqref{28}, we get, 
\begin{equation}
S = p_1(b_0(1-r) - b_1) + p_2(b_1(1-r) - b_2) + \cdots + p_N(b_{N-1}(1-r) - b_N) \geq 0.
\label{29}
\end{equation}
Rearranging Eq. \eqref{29} by collecting the storage terms together and setting $b_0$ and $b_N$ to zero, we have, 
\begin{equation}
b_1(p_2(1-r) - p_1) + b_2(p_3(1-r) - p_2) + \cdots + b_{N-1}(p_N(1-r) - p_{N-1}) \geq 0.
\label{30}
\end{equation}
In order to maximize savings, storage will not take place when the individual terms on the left hand side of Eq. \eqref{30} is negative. Since $b_i$ represents the state of battery storage at the end of period $i$ which is always positive, the maximum savings from storage will occur when, 
\begin{equation}
p_2(1-r) - p_1 \geq 0, p_3(1-r) - p_2 \geq 0, \cdots, p_N(1-r) - p_{N-1} \geq 0. 
\label{31}
\end{equation}
Thus, from Eq. \eqref{31}, for a two-period model, storage during period 1 will be beneficial during period 2 only if,
\begin{equation}
\frac{p_1}{p_2} \leq (1-r). 
\label{32}
\end{equation}
Similarly, for a three-period model, storage in period 1 will be beneficial during period 2 only if $p_1 \leq p_2(1-r)$ and storage in period 2 will be beneficial during period 3 only if $p_1 \leq p_2(1-r)$. Thus, storage in period 1 will be beneficial during period period 3 only if,
\begin{equation}
\frac{p_1}{p_3} \leq (1-r)^2.
\label{33}
\end{equation}
Generalizing to a $N$-period model, storage in any period $i$ will be beneficial during period $i+k$ , if and only if,
\begin{equation}
\frac{p_i}{p_{i+k}} \leq (1-r)^k,  i \in {1, 2, \cdots, N-1}, k \in {1, 2, \cdots, N-i}.
\label{34}
\end{equation}
In other words, storage in any period $i$ is maximally beneficial during period $i+k$ only if the ratio of prices at period $i$ to that of period $i+k$ is less than the storage efficiency $(1-r)$ of the battery over $k$ periods.

\bibliographystyle{IEEEtran}
\bibliography{9_Ref}

\end{document}